\documentstyle[11pt,epsf]{article}

\def\fnote#1#2{\begingroup\def\thefootnote{#1}\footnote{#2}\addtocounter
{footnote}{-1}\endgroup}

\renewcommand{\thefootnote}{\fnsymbol{footnote}}

\def\beq{\begin{eqnarray}}
\def\eeq{\end{eqnarray}}
\def\bea{\begin{eqnarray*}}
\def\eea{\end{eqnarray*}}
\def\gev{\, {\rm GeV}}

\def\NPB#1#2#3{Nucl. Phys. {\bf B#1}, #3 (#2)}
\def\PLB#1#2#3{Phys. Lett. {\bf B#1}, #3 (#2)}
\def\PLBold#1#2#3{Phys. Lett. {\bf #1B}, #3 (#2)}

\def\PRD#1#2#3{Phys. Rev. {\bf D#1}, #3 (#2)}

\def\PRL#1#2#3{Phys. Rev. Lett. {\bf #1}, #3 (#2)}
\def\PREP#1#2#3{Phys. Rep. {\bf #1}, #3 (#2)}

\def\centeron#1#2{{\setbox0=\hbox{#1}\setbox1=\hbox{#2}\ifdim
\wd1>\wd0\kern.5\wd1\kern-.5\wd0\fi
\copy0\kern-.5\wd0\kern-.5\wd1\copy1\ifdim\wd0>\wd1
\kern.5\wd0\kern-.5\wd1\fi}}
\def\ltap{\;\centeron{\raise.35ex\hbox{$<$}}{\lower.65ex\hbox{$\sim$}}\;}
\def\gtap{\;\centeron{\raise.35ex\hbox{$>$}}{\lower.65ex\hbox{$\sim$}}\;}
\def\gsim{\mathrel{\gtap}}
\def\lsim{\mathrel{\ltap}}

\def\slashchar#1{\setbox0=\hbox{$#1$}           
   \dimen0=\wd0                                 
   \setbox1=\hbox{/} \dimen1=\wd1               
   \ifdim\dimen0>\dimen1                        
      \rlap{\hbox to \dimen0{\hfil/\hfil}}      
      #1                                        
   \else                                        
      \rlap{\hbox to \dimen1{\hfil$#1$\hfil}}   
      /                                         
   \fi}                                        %

\def\singleandthirdspaced{\baselineskip=\normalbaselineskip\multiply
    \baselineskip by 130\divide\baselineskip by 100}
\def\singlespaced{\baselineskip=\normalbaselineskip}

\newcommand{\newc}{\newcommand}
\newc{\Tr}{{\rm tr}}
\newc{\GG}{{\stilde G}}
\newc{\slepton}{{\widetilde\ell}}
\newc{\Det}{{\rm det}}
\newc{\Lambdaprime}{{\Lambda}}
\newc{\LambdaQCD}{\Lambda_{\rm eff}}
\newc{\mrat}{{\tilde m}}
\newc{\mratprime}{{\tilde m^\prime}}
\newc{\Log}{{\rm log}}
\newc{\Nc}{{N_c}}
\newc{\Nmess}{{N_{\rm mess}}}
\newc{\Mmess}{{M_{\rm mess}}}
\newc{\stilde}{\widetilde}
\newc{\stau}{{\stilde\tau}}
\newc{\stauI}{{\stilde\tau_1}}
\newc{\HIT}{{HIT}}
\newc{\Nf}{{N_f}}
\newc{\aT}{{a_T}}
\newc{\aB}{{a_B}}
\newc{\bT}{{a_m}}
\newc{\dT}{{a_{M}}}
\newc{\cT}{c_T}
\newc{\mQ}{m_{\tilde Q}}
\newc{\mg}{m_{\tilde g}}
\newc{\msoft}{m_{\rm soft}}

\begin{document}

\begin{titlepage}
\begin{flushright}
{\large
hep-ph/9805289 \\
SLAC-PUB-7827\\
May 11, 1998
}
\end{flushright}

\vskip 1.2cm

\begin{center}

{\LARGE\bf Cornering gauge-mediated supersymmetry breaking}

{\LARGE\bf with quasi-stable sleptons at the Tevatron}

\vskip 2cm

{\large
 Stephen P.~Martin$^a$
and James D.~Wells$^b$\fnote{\dagger}{Work supported by the Department of
Energy under contract DE-AC03-76SF00515. } 
} \\
\vskip 4pt
{\it $^a$Physics Department,
     University of California,
     Santa Cruz CA 95064  } \\
{\it $^b$Stanford Linear Accelerator Center,
     Stanford CA 94309} \\

\vskip 1.5cm

\begin{abstract}

There are many theoretical reasons why heavy quasi-stable charged
particles might exist.  Pair production of such particles at the Tevatron
can produce highly ionizing tracks (HITs) or fake muons.  In
gauge-mediated supersymmetry breaking, sparticle production can lead to
events with a pair of quasi-stable sleptons, a significant fraction of
which will have the same electric charge. Depending on the production
mechanism and the decay chain, they may also be accompanied by additional
energetic leptons. We study the relative importance of the resulting
signals for the Tevatron Run II.  The relative fraction of same-sign
tracks to other background-free signals is an important diagnostic tool in
gauge-mediated supersymmetry breaking that may provide information about
mass splittings, $\tan\beta$ and the number of messengers communicating
supersymmetry breaking. 

\end{abstract}

\end{center}

\vskip 1.0 cm

\end{titlepage}
\setcounter{footnote}{0}
\setcounter{page}{2}
\setcounter{section}{0}
\setcounter{subsection}{0}
\setcounter{subsubsection}{0}

\singleandthirdspaced

\indent

Low-energy supersymmetry has emerged as an excellent candidate solution to
the hierarchy problem associated with the 
existence of the small ratio $M_W/M_{\rm Planck}$ in the
Standard Model. 
However, the existence of squarks and sleptons
in the Minimal Supersymmetric Standard Model (MSSM) seems to lead
to another potential difficulty: the supersymmetric flavor problem. 
If the soft supersymmetry-breaking mass parameters for
squarks, sleptons, and gauginos do not greatly exceed 1 TeV, as suggested
by a solution to the hierarchy problem, then arbitrary mixing angles
associated with these mass terms can induce unacceptably
large flavor-changing effects in
low-energy processes like $\mu \rightarrow e \gamma$,
$K^0 \leftrightarrow \overline K^0$, $b \rightarrow s \gamma$, etc.
Conversely, the required
absence of such flavor violation can be viewed as a strong
constraint, and therefore a powerful clue, regarding the nature of
supersymmetry breaking.

Historically, the more popular approach has been to assume that 
supersymmetry-breaking effects have their origin in a ``hidden sector",
and are then communicated only (or dominantly) by Planck-suppressed
effects to the fields of the MSSM. However, within this
framework, the absence of low-energy flavor-changing neutral currents
really depends on further implicit assumptions, since {\it a priori} it
is just as likely to mediate supersymmetry breaking to the MSSM with
flavor-violating Planck-suppressed operators as with flavor-blind ones.
One may impose
an approximate flavor symmetry on the relevant terms in the lagrangian,
but it is rather controversial whether this can be well-motivated
theoretically by deeper principles.

An alternative hypothesis 
\cite{oldGMSBmodels,newGMSBmodels} is that the ordinary gauge interaction
$
SU(3)_C \times SU(2)_L \times U(1)_Y
$
are responsible for communicating supersymmetry breaking effects to the
MSSM fields. In these gauge-mediated supersymmetry-breaking (GMSB) models,
the absence of large flavor-violating effects
in low-energy physics is a natural consequence of the flavor-blindness
of the Standard Model gauge interactions. The supersymmetry-breaking
sector of the theory couples to some new ``messenger" quark
and lepton superfields with vector-like $
SU(3)_C \times SU(2)_L \times U(1)_Y
$ interactions. For example, suppose that the messenger quarks and leptons
come in $\Nmess$ copies of the ${\bf 5} +
{\bf \overline 5}$ representation of the global $SU(5)$ symmetry which
includes $            
SU(3)_C \times SU(2)_L \times U(1)_Y
$. 
Because of the effects of dynamical supersymmetry breaking, there
is a small splitting among the messenger fermion and scalar masses.
In the simplest types of models with only one $F$-term supersymmetry
breaking order parameter, this
can be parameterized as follows.
For each messenger
fermion with mass $m_{\psi_i}$, the messenger scalar partner masses
are given by $m_{\psi_i} \sqrt{1 \pm \Lambda/m_{\psi_i} }$, 
where $\Lambda$ is a constant mass scale
which is the same for all of the messenger supermultiplets.
In order for the messenger scalars not to develop color- or
charge-breaking vacuum expectation values, it is necessary  
that $\Lambda < m_{\psi_i}$ for each messenger supermultiplet.
It is usual to assume that the messenger particles are roughly
degenerate, with masses all of order $\Mmess$, so that
$\Lambda$ can be treated as a perturbation with respect to
$\Mmess \approx m_{\psi_i}$. 

With these assumptions, the MSSM gaugino and scalar soft
supersymmetry-breaking masses can be easily calculated 
at leading order in an expansion in $\Lambda/\Mmess$
\cite{newGMSBmodels}. Gaugino masses
are communicated from the messenger sector to the MSSM
at one loop
\beq
M_a = \Nmess \Lambda {\alpha_a \over 4\pi} \qquad\qquad
(a=1,2,3),
\label{gauginomasses}
\eeq
and are proportional to the corresponding squared gauge couplings.
Squark, slepton, and Higgs boson squared masses arise at two
loops and are given by
\beq
m_\phi^2 = 2 \Nmess \Lambda^2 \sum_{a=1}^3 \left ({\alpha_a \over 4 \pi}
\right
)^2 C_a^\phi
\label{scalarmasses}
\eeq
where $C_3^\phi$ is equal to $4/3$ for each squark and 0
for other scalars; $C_2^\phi$ is equal to $3/4$ for weak isodoublet
scalars and $0$ for weak isosinglets, and $C_1^\phi = 3Y_\phi^2/5$
for each scalar of weak hypercharge $Y_\phi$ with $\alpha_1$ in a GUT
normalization.
Equations
(\ref{gauginomasses}) and (\ref{scalarmasses}) are subject to
corrections in $\Lambda/\Mmess$ 
which turn out to be usually quite small \cite{grat} and will be neglected
in the following. The sparticle spectrum can now be computed
by using renormalization group equations to run the masses  
eqs.~(\ref{gauginomasses}) and (\ref{scalarmasses}) and 
other couplings from the scale $\Mmess$ down to the electroweak
scale~\cite{peskin,DDRT,KKW}. 
This class of models is therefore highly predictive,
with input parameters $\Lambda$, $\Mmess$, 
$\Nmess$, $\tan\beta$, and sgn$(\mu)$, and the phenomenology
is quite distinctive~\cite{DDRT}-\cite{lep}.

The prediction of a goldstino/gravitino
($\GG$)~\cite{Fayet,oldphotonsignals} 
as the lightest supersymmetric particle (LSP) is another general feature
of GMSB models. 
In terms of the parameters $\Lambda$ and $\Mmess$ above,
the supersymmetry-breaking order parameter is $\langle F \rangle
= C \Lambda \Mmess$, where $C$ is a dimensionless constant which
can be of order unity (for ``direct" gauge-mediation models), or
much larger than 1 (for ``indirect" gauge-mediation models),
but not much less than 1.
Each of the MSSM sparticles can decay into final states including the
goldstino/gravitino $\GG$, with rates proportional
to $1/\langle F \rangle^2$. 
However, the decays of MSSM sparticles to the goldstino will typically 
be dominated by other kinematically-allowed decays, except in the case of
the next-to-lightest
supersymmetric particle (NLSP). If $R$-parity is conserved, as motivated
by the absence of rapid proton
decay, then the NLSP can only decay to 
its Standard Model partner(s) and the goldstino/gravitino.
Whether the NLSP can decay quickly
enough to be visible within a collider detector depends on the identity
and mass of the NLSP, and on the goldstino decay constant 
$\langle F \rangle$. 
If $\langle F \rangle$ is less than a
few thousand TeV, then one can hope to observe decays to the
goldstino within a typical collider detector, with potentially
spectacular consequences. Conversely, if $\langle F \rangle \gg
10^3$ TeV, then all decays involving $\GG$ will occur far outside
the detector.

In GMSB models of the type discussed above, 
the NLSP is generally either the
lightest neutralino ($\stilde N_1$)
or a charged slepton, depending on the model parameters. 
In this paper, we will concentrate on the latter case.
The three lightest sleptons generally consist of the nearly unmixed
and degenerate right-handed selectron and smuon 
$\stilde e_R$ and $\stilde \mu_R$, and
a mixed stau mass eigenstate $\stilde \tau_1$. The reason for this
is that each of the slepton (mass)$^2$ matrices contains an off-diagonal
term $-\mu m_\ell \tan\beta$ where $m_\ell$ is the mass of the
corresponding lepton. This
provides for slepton mixing and
lowers the corresponding slepton mass eigenvalue. In the case of the
selectron and smuon, this effect is quite small, only reducing the
mass of the smuon by at most a few tens of MeV, and the 
mass of the selectron by
much less. Therefore we will simply neglect smuon and selectron
mixing and treat them as degenerate, unmixed states. However, because of
the
hierarchy 
$m_\tau \gg m_\mu$, the stau mixing is not negligible unless $\tan\beta$
is close to 1,
so that 
\beq
m_{\stilde \tau_1} < m_{\stilde e_R} \simeq m_{\stilde
\mu_R}.
\label{ordering}
\eeq

Therefore, it is useful to
distinguish between two qualitatively distinct scenarios, depending on
whether or not the
right-handed selectrons and smuons ($\stilde e_R$ and $\stilde \mu_R$) can
have kinematically allowed decays
into the lightest stau $\stilde \tau_1$. 
If the mass difference
$m_{\stilde \ell_R} - m_{\stilde \tau_1}$ exceeds about 1.8 GeV, then
one can have three-body decays 
\beq
\stilde \ell_R^- \rightarrow \ell^- \tau^\pm \stilde \tau_1^\mp 
\label{threebody}
\eeq
for $\ell = e$ or $\mu$, and similarly for $\stilde\ell_R^+$.
In that case we have a ``stau NLSP scenario", in which
all supersymmetric decay chains end up in $\stilde \tau_1$, with
a subsequent (possibly very slow) decay 
\beq
\stilde \tau_1 \rightarrow \tau
\GG .
\label{staugoldstinodecay}
\eeq
Conversely, if $\stilde \ell_R$ and $\stilde \tau_1$ are degenerate in
mass to within less than 1.8 GeV, then the aforementioned three-body
decays are not kinematically allowed. In this ``slepton co-NLSP" scenario,
the three sleptons $\stilde e_R$, $\stilde \mu_R$ and $\stilde \tau_1$
each act {\it effectively} as the NLSP
despite eq.~(\ref{ordering}), in the sense that they only
have
kinematically allowed decays into the goldstino.\footnote{An exception
occurs if $|m_{\stilde N_1} - m_{\stilde \tau_1}| < m_{\tau}$ and
$m_{\stilde \ell_R} > m_{\stilde N_1}$, which corresponds to a
neutralino-stau co-NLSP scenario.}  
The lightest stau
will decay according to 
eq.~(\ref{staugoldstinodecay}), while the lightest selectron and
smuon decay according to 
$\stilde e_R \rightarrow e \GG $ and 
$\stilde \mu_R \rightarrow \mu \GG $ respectively.

In this paper, we will consider slepton co-NLSP models and stau NLSP
models, with the subsequent decays to the goldstino/gravitino $\GG$ 
assumed to be very slow, so that they always occur outside the
detector. (If instead those decays occur promptly, or with a macroscopic
decay length but within the detector, then the signals from
additional hard leptons and/or decay kinks or impact parameters will
be even more spectacular.) The quasi-stable sleptons arising from
supersymmetric events can then manifest themselves in different ways in a
detector, depending on how fast they are~\cite{DreesTata,stuart,FM}. 
The relevant kinematic
variable is
$\beta\gamma = 
\left (E^2/m_{\stilde \ell}^2 - 1 \right )^{1/2}$
where $E$ is the relativistic energy of the slepton in the lab frame.
For $\beta\gamma \gsim 1$ or so, the ionization rate $-dE/dx$ of the
slepton as it moves through the detector material is minimal, and the 
fast slepton penetrates the detector, mimicking a ``muon". Slow sleptons
with $\beta\gamma \lsim 1$ have a greater-than-minimum ionization rate as
they move through the detector material. The ionization rate increases
sharply as $\beta\gamma$ decreases, so that for $\beta\gamma < 0.85$ or
$0.9$ the resulting Highly Ionizing Track (HIT) can be readily
distinguished from that of a muon~\cite{DreesTata,stuart,DDRT,FM}.

At the Tevatron Run II, the most important
sparticle production mechanisms are then typically slepton production,
\beq
p\overline p \rightarrow \stilde e_R^+ \stilde e_R^-,\>
\stilde \mu_R^+ \stilde \mu_R^-\,\>\,{\rm or}\>\,\, \stilde \tau_1^+ \stilde
\tau_1^-
\label{sleptonproduction}
\eeq
and/or chargino/neutralino production,
\beq
p\overline p &\rightarrow & \stilde C_1^+ \stilde C_1^-\, \>{\rm or}\>
\,
\stilde C_1^\pm \stilde N_2 .
\eeq
Of course, other processes can contribute small amounts to
the signal.
Production of heavier slepton pairs ($\stilde \nu \stilde \nu$,
$\stilde \nu \stilde \ell^{\pm}_L$, $\stilde \ell^{+}_L\stilde
\ell^{-}_L$)
is generally less important than eq.~(\ref{sleptonproduction}), but 
may still be observable.
Other chargino
and neutralino combinations ($\stilde C_i^+ \stilde C_j^-,$
$\stilde C_i^\pm \stilde N_j$, and $\stilde N_i \stilde N_j$)
might give significant contributions, especially if the higgsino contents
of
the $\stilde N_1$, $\stilde N_2$ and $\stilde C_1$ are not negligible.
Production of gluinos and squarks is generally quite negligible
within the class of GMSB models we consider, because they
tend to be relatively heavy.
In the simulations described below we simply include all contributions
to sparticle pair production.

Each supersymmetric event leads to a pair of quasi-stable sleptons which
may be identified as either a ``muon" or a HIT. In addition,
a high percentage of these events can actually have quasi-stable sleptons
with the same charge in the final state. For example, in the case of
$\stilde C_1^\pm
\stilde N_2$ production, the $\stilde N_2$ will decay  equally to sleptons
with either charge. Events with $\stilde C_1^+ \stilde C_1^-$ 
(or $\stilde \nu \stilde \nu$,
$\stilde \nu \stilde \ell^{\pm}_L$, $\stilde \ell^+_L\stilde \ell^-_L$)
production
will lead to roughly equal numbers of like- and opposite-charge slepton
NLSPs whenever any part of the decay chains involves a real or virtual
neutralino, because the neutralinos are Majorana particles and do not
know about electric charge. In the slepton co-NLSP scenario,
$\stilde e_R^+ \stilde e_R^-$,
$\stilde \mu_R^+ \stilde \mu_R^-$, and 
$\stilde \tau_1^+ \stilde \tau_1^-$
production always leads to opposite-charge sleptons.
However, in the stau-NLSP scenario, 
$\stilde e_R^+ \stilde e_R^-$ and
$\stilde \mu_R^+ \stilde \mu_R^-$ production usually leads to roughly
equal numbers of same-sign and opposite-sign staus in the final state.
This is because the three-body decays 
$\stilde \ell_R^- \rightarrow \ell^- \tau^\pm \stilde \tau_1^\mp
$
go through a virtual neutralino, so that the charge of the $\stilde
\tau_1$
is nearly uncorrelated with the charge of its parent $\stilde \ell_R$
when $m_{\stilde N_1}$ is not much larger than $m_{\stilde \ell_R}$.
As $m_{\stilde N_1}/m_{\stilde \ell_R}$ increases, the branching
fraction
for the ``slepton charge-flipping'' decays
$\stilde \ell_R^- \rightarrow \ell^- \tau^- \stilde \tau_1^+ $
increases at the expense of the ``slepton charge-preserving" decays
$\stilde \ell_R^- \rightarrow \ell^- \tau^+ 
\stilde \tau_1^- $~\cite{threebodyakm}.
However, for reasonable values of $m_{\stilde N_1}/m_{\stilde \ell_R}$
found in GMSB models with a $\stilde \tau_1$ NLSP 
as studied here, the number of events 
from $\stilde e_R^+ \stilde e_R^-$ and
$\stilde \mu_R^+ \stilde \mu_R^-$ production with
like-sign staus will be comparable to (albeit smaller than)
the number with opposite sign staus.
If, as is often the case in models, the mass difference
$m_{\stilde \ell_R} - m_{\stilde\tau_1}$ is not too large,
then the $\ell$ and the $\tau$ produced in the decay will be very soft
and will fail to pass cuts for lepton identification.
 
In order to define our signals, we require that events pass 
at least one of the
following two triggers.\footnote{We are grateful
to D.~Stuart for explaining the trigger and \HIT~identification
requirements relevant for CDF in Run II.} 
First, events are triggered if at least one quasi-stable
slepton
has $|\eta| < 0.6$ and $\beta\gamma > 0.4$. 
The pseudorapidity
requirement corresponds to the highly-shielded central region of the CDF
detector in order to cut down on backgrounds.  
The lower limit on 
$\beta\gamma $ ensures that the slepton will penetrate the calorimeters.
Second, events are triggered with at least one fast
quasi-stable slepton which mimics a high-$p_T$ central muon.
This requires the trigger slepton to satisfy $\beta\gamma >
0.85$ and
$|\eta| < 1.0$. The lower limit on $\beta\gamma$ is to ensure that the
fast slepton will have a reasonable probability of penetrating the
detector within a narrow time window in order to satisfy
identification requirements for a muon.
(Note that if a slepton with mass greater than 90 GeV satisfies these
requirements, it will necessarily have $p_T > 50$ GeV in the case of the
first trigger and $p_T > 30$ GeV in the case of the second trigger,
so we do not require a separate $p_T$ cut.) We find that in most of
the cases
studied below,
the percentage of supersymmetric events which pass at least one of these
two trigger
requirements is quite high, typically between 70\% and 85\%.

After an event comes in on trigger, we identify particles according
to the following criteria:
\begin{itemize}
\item[$\bullet$] A quasi-stable slepton is identified as a \HIT~if it has
$|\eta| < 1.0$ and $0.4 < \beta\gamma < 0.85$  and $p_T > 30$ GeV. 
\item[$\bullet$] A quasi-stable slepton is identified as a ``muon" if it
fails the HIT requirement and satisfies
$|\eta| < 1.7$ and $\beta\gamma > 0.85$. 
\item[$\bullet$] A real electron or muon must satisfy 
$|\eta| < 1.7$
and 
$p_T > 12$ GeV. 
\item[$\bullet$] A jet must satisfy 
$|\eta | < 3.0$
and
$p_T > 15$ GeV. 
\end{itemize}
In the case of a \HIT, ``muon" or a real lepton, we impose an isolation
requirement that within a cone $\sqrt{(\Delta \eta)^2 +
(\Delta \phi)^2} < 0.4$ there should be no other \HIT, ``muon" or lepton
and that the total hadronic energy should not exceed 5 GeV.

Within this trigger sample, we now define the following signals:
\begin{itemize}
\item[1)] \HIT: Events with at least one isolated slepton identified
as a \HIT. 
\item[2)] SS: A pair of {\it same-sign} fast sleptons each passing the
``muon"
cut
above, with no other isolated leptons.
\item[3)] 3$\ell$: Trilepton signal consisting of two fast sleptons which
each pass the ``muon" cut
above, and exactly one additional isolated $e$ or $\mu$, and no jets.
\item[4)] $4\ell +$: 
Four or more isolated lepton candidates, including two
fast sleptons which each pass the ``muon" cut above.
\end{itemize}
In the trilepton signal case $(3\ell)$, we demand that no pair of
oppositely
charged muon candidates reconstructs to an invariant mass $m_Z\pm 10$
GeV in order
to reduce the backgrounds from $WZ$ production. (Here we use 
the invariant mass as reconstructed from the 3-momenta of the particles
assuming they are essentially massless, which does not coincide with the
true invariant mass since at least one of the pair is actually a massive
slepton.) We also do not
allow any jets in the event from initial state radiation or primary
decay products. Such a jet veto avoids potentially large backgrounds from
$t\bar t$ production.  In the $4l+$ signal we require that it not be
consistent with $ZZ$ production.  The invariant mass cuts to accomplish
this are the same as for the $3l$ signal discussed above.
We should also note that 
a significant fraction of the $4l+$
events will have 5 or 6 isolated leptons arising from gaugino cascade
decays.
Since the ``muons" in these events always have $p_T > 50$ GeV
for $m_{\slepton} > 90$ GeV, there should be essentially no background
after cuts for all of the signals proposed above, except the $3l$
signal whose background can be rendered insignificant with sufficiently
hard $p_T$ cuts on the ``muons.''

Let us now study the relative importance of the
signals defined above 
for $\overline p p$ collisions at $\sqrt{s} = 2$ TeV, as is 
relevant for the Tevatron Run II scheduled to begin in 2000. 
We will examine representative models in the parameterization
of eqs.~(\ref{gauginomasses})-(\ref{scalarmasses}).  
Our collider
simulations have been performed using ISAJET \cite{isajet}.
A first
quantitative study 
with somewhat different emphasis and different definitions of
signals and cuts
has been carried out in ref.~\cite{FM}. 

\begin{figure}[tbh]
\centering
\epsfxsize=4.1in
\hspace*{0in}
\epsffile{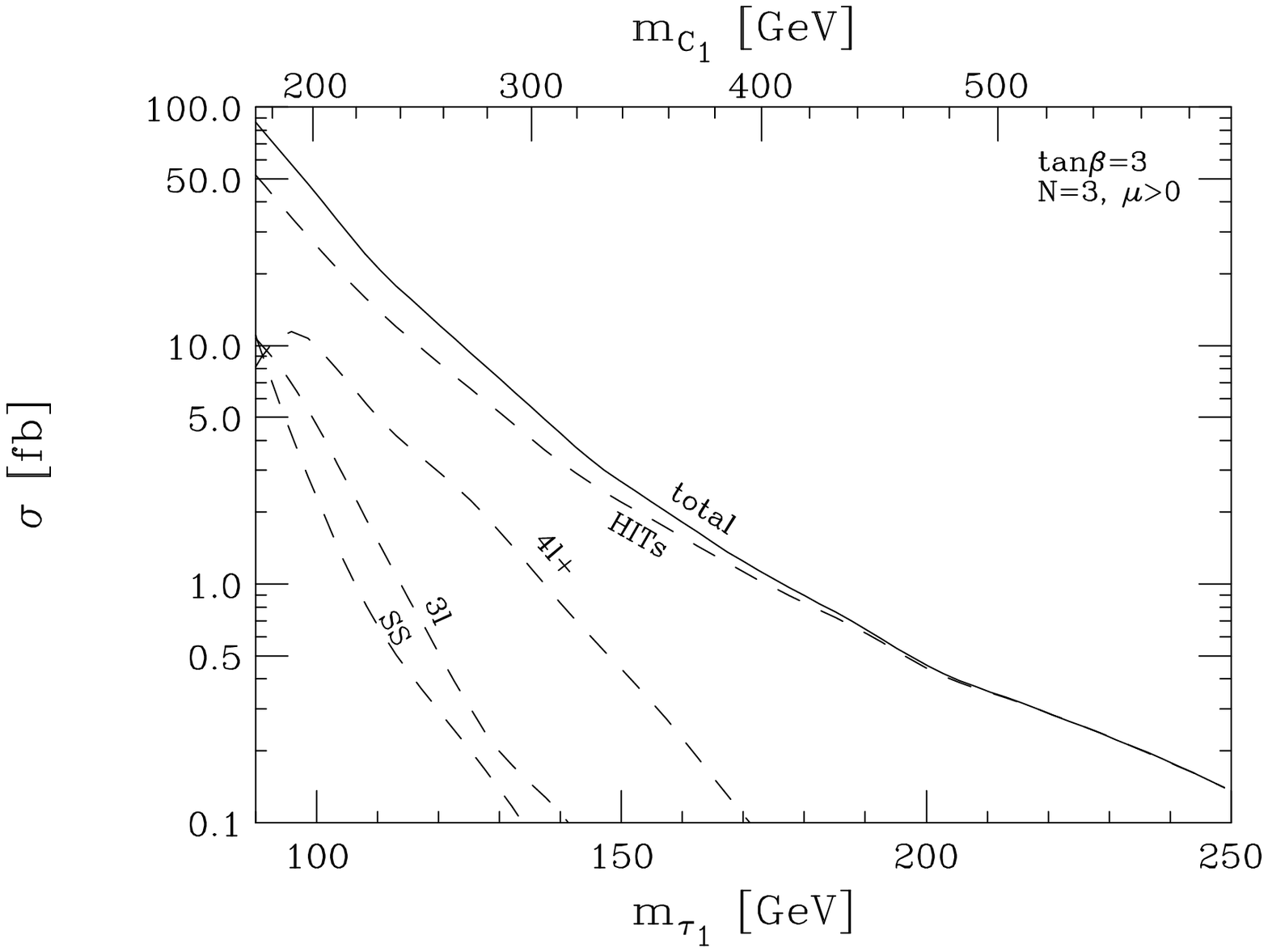}
\caption{Cross-sections in fb for producing various final
states from sparticle pair-production in $p\overline p$ collisions at
$\sqrt{s} = 2$ TeV. The dashed lines are rates for events with 
at least one slepton identified as a highly ionizing track (HITs); 
four or more leptons including one or more fast sleptons
masquerading as ``muons" ($4\ell+$); 
trileptons including one or more ``muons" ($3\ell$); and
same-sign ``muons" with no other isolated leptons (SS). 
The results shown are for slepton co-NLSP GMSB models with varying
$\Lambda = \Mmess/3$, and
fixed $\Nmess = 3$, $\tan\beta = 3$, and $\mu > 0$.
The solid line is the sum of the four signals.
}
\label{tb3varylambda}
\vspace*{0.3in}
\epsfxsize=4.1in
\hspace*{0in}
\epsffile{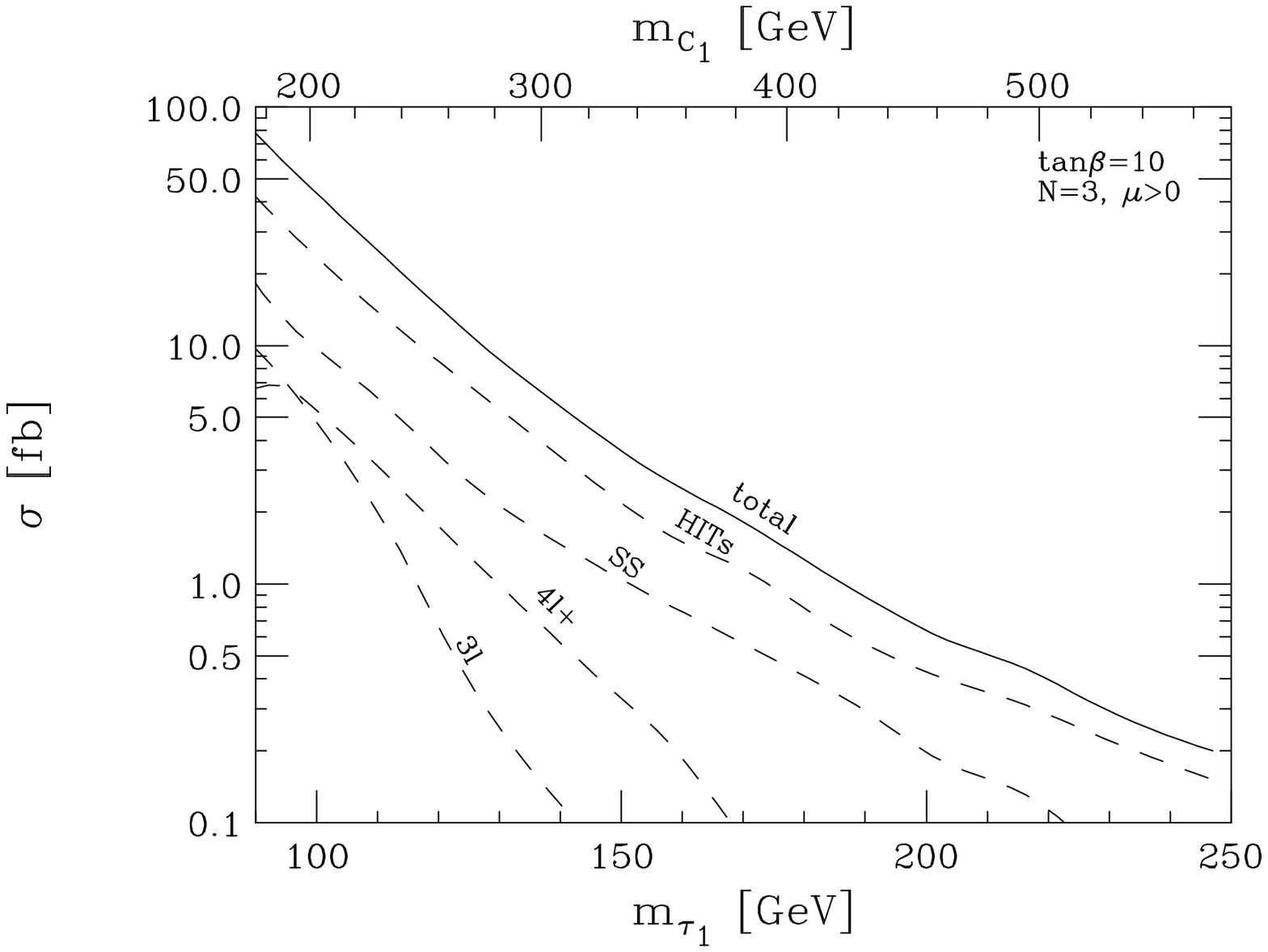}
\caption{As in Figure \ref{tb3varylambda}, but for stau NLSP
models with $\tan\beta= 10$.
Note that the component of the signal due to same-sign ``muons" is much
more significant
in this case.
}
\label{tb10varylambda}
\end{figure}

We first consider a one-parameter family of slepton
co-NLSP models with $\Nmess = 3$, $\tan\beta = 3$, $\mu > 0$
(in the sign convention of ref.~\cite{conventions}) and
varying $\Lambda$ with $\Mmess = 3\Lambda$. 
(The factor of 3 here is rather arbitrary, but the sparticle spectrum
depends on $\Mmess$ only logarithmically anyway.)
Since the LEP2 collaborations \cite{lep} should be able to
rule out slepton masses up to at least 90 GeV in the scenarios we
consider, we take $\Lambda$ to vary over a range 
27 TeV $<\Lambda <$ 80 TeV which corresponds to 
90 GeV $< m_{\stilde \tau_1} <$ 250 GeV. In this family of models,
the mass differences $m_{\stilde e_R} - m_{\stilde \tau_1}$ 
and $m_{\stilde \mu_R} - m_{\stilde \tau_1}$ are
always positive and less than 1 GeV, so that the three-body decays
in eq.~(\ref{threebody}) are not open, and $\stilde
e_R$, $\stilde
\mu_R$,
and $\stilde \tau_1$ are effectively co-NLSPs.
We note that in the lower mass regions ($m_{\stilde \tau_1}\lsim 150\gev$)
the
chargino and neutralino production rate constitutes the largest
source of supersymmetry events at the Tevatron.  At higher
mass regions ($m_{\stilde \tau_1}\gsim 150\gev$) it is slepton
production which dominates.
In Fig.~\ref{tb3varylambda} we show the
four signal cross-sections and their total after the trigger,
identification and
isolation cuts described above. 
Using a discovery criterion of 7 total signal events, we find
that a discovery should be possible for $m_{\stau_1} < $(140, 185, 225)
GeV in
these models for
an integrated luminosity of (2, 10, 30) fb$^{-1}$.
These $\stilde \tau_1$ mass limits correspond to limits
on the lightest chargino mass, $m_{\stilde C_1}<(320,430,540)$ GeV in
these particular models, but it is important to note that slepton
production
is the dominant contribution to the signal for larger masses. 
The \HIT~signal is clearly the largest single 
one over the entire range, but the $4\ell +$ signal can be a significant
component, and for smaller values of $m_{\stau_1}$ the $3\ell$ and
SS signals can also be observed with sufficient integrated luminosity. 
However, for larger sparticle masses, the discovery signal comes
almost entirely from the \HIT~signal.  

Next we consider a similar one-parameter family of models with all other
parameters as before,
but now
with $\tan\beta = 10$. Because of the larger mixing in the
stau (mass)$^2$ matrix, the mass
differences 
$m_{\stilde e_R} - m_{\stilde \tau_1}$    
and $m_{\stilde \mu_R} - m_{\stilde \tau_1}$ are
now always greater than 3 GeV for $m_{\stilde \tau_1} < 250$ GeV, so that
the
three-body decays
$\stilde e_R \rightarrow e \tau \stau_1$ and
$\stilde \mu_R \rightarrow \mu \tau \stau_1$ are kinematically allowed.
These models are therefore examples of the stau NLSP scenario. The
corresponding 
signal
cross-sections are shown in Fig.~\ref{tb10varylambda}. Again, the
\HIT~signal is the largest one, but the SS signal is quite significant
over the entire range, amounting to about 20 to 35\% 
of the total signal events.
Most of these SS events arise from direct
production of 
$\stilde e_R^+ \stilde e_R^-$ or $\stilde \mu_R^+ \stilde \mu_R^-$
with subsequent three-body decays. 
The existence of the SS signal allows the discovery reach in $m_{\stilde
\tau_1}$ to be slightly higher (about 10 GeV) in these stau NLSP models
than in the analogous slepton co-NLSP models 
in Fig.~\ref{tb3varylambda}.
(In slepton co-NLSP models,
direct 
$\stilde e_R^+ \stilde e_R^-$,
$\stilde \mu_R^+ \stilde \mu_R^-$, and
$\stilde \tau_1^+ \stilde \tau_1^-$ production
can never lead to a SS signal.)
The $4\ell +$ signal can also be important for smaller values of
$m_{\stilde \tau_1}$.

\begin{figure}[tbh]
\centering
\epsfxsize=4.8in
\hspace*{0in}
\epsffile{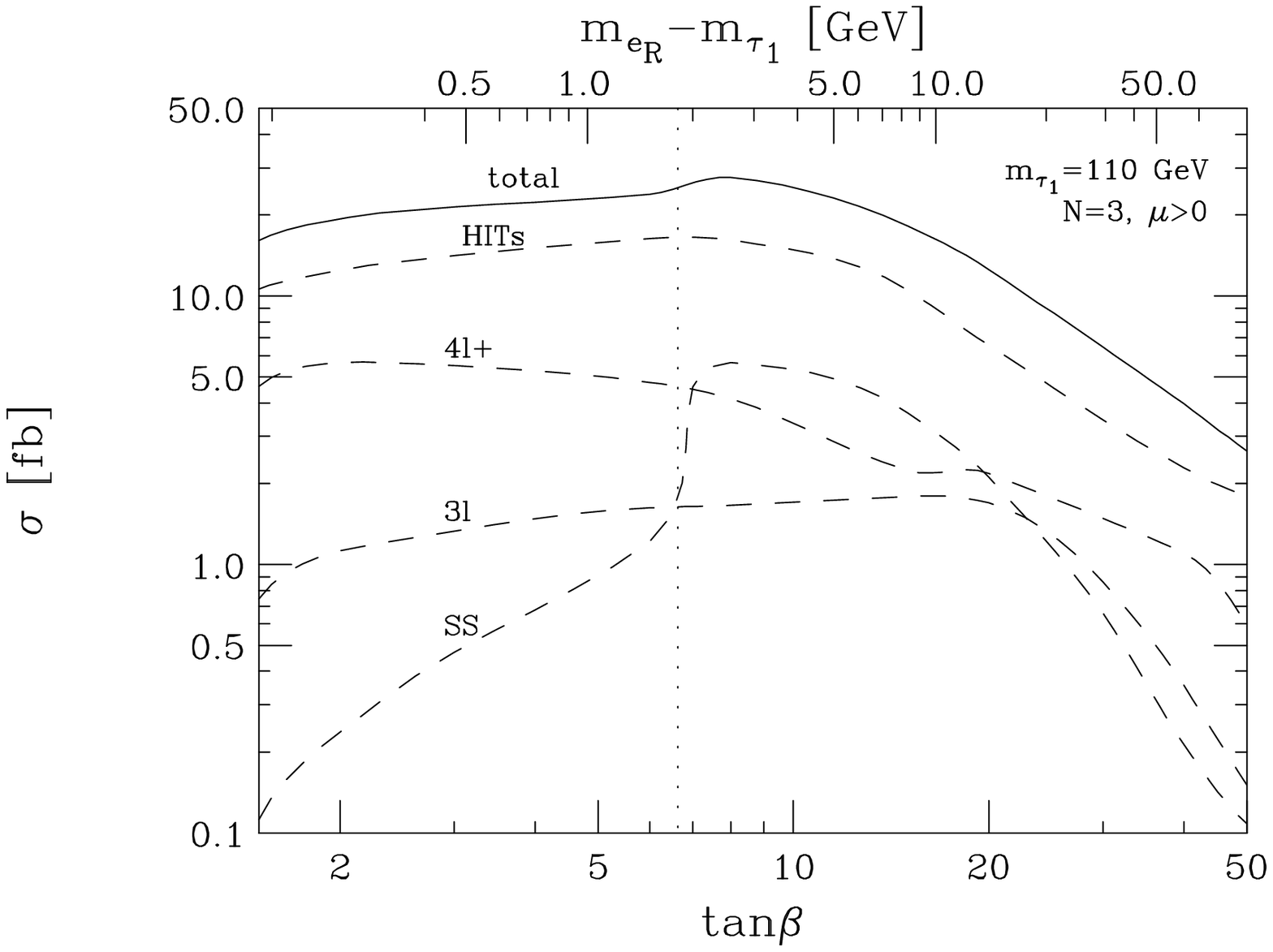}
\caption{Cross-sections in fb for producing various final
states from sparticle pair-production in $p\overline p$ collisions at
$\sqrt{s} = 2$ TeV, labelled as in Figures 1 and 2. 
The results shown are for GMSB models with varying
$\tan\beta$, and with $\Lambda = \Mmess/3$ chosen so that
$m_{\stilde \tau_1}$ is fixed at 110 GeV, with $\Nmess = 3$, and $\mu >
0$.
The dotted vertical line is the nominal boundary between the slepton co-NLSP
scenario (on the left) and the stau NLSP scenario (on the right).}
\label{varytb}
\end{figure}

In Fig.~\ref{varytb} we show the same signals but now with varying
$\tan\beta$,
with $N_{\rm mess}=3$,
$\mu > 0$, and $\Lambda = \Mmess/3$ chosen so that
$m_{\stilde \tau_1}$ is fixed at 110 GeV.  
The mass difference between
$\stilde e_R$ (or equivalently, $\stilde \mu_R$) and
$\stilde \tau_1$ is followed on the upper horizontal axis.  
For $\tan\beta \gsim 6.6$ the three-body decays for $\stilde e_R$ and
$\stilde \mu_R$ open up, and the
same-sign ``muon''
signal becomes large.  
\footnote{There is a small range of $\tan\beta$ near the boundary between
the two scenarios
for which the mass differences $m_{\stilde \ell_R} - m_\ell - m_{\tau}
- m_{\stilde \tau_1}$ are positive but small, so that the three-body
decays can have a macroscopic decay
length \cite{threebodyakm}. We will not explore that interesting
possibility in
the
following.} 
For very large values of $\tan\beta$, the
masses of the $\stilde e_R$ and $\stilde \mu_R$ 
must be much higher than the lighter
stau mass, which is fixed at 110 GeV for the figure.  
Therefore,
the largest source for SS dimuons -- $\stilde e_R^-\stilde e_R^+$
and $\stilde \mu_R^-\stilde \mu_R^+$ -- gets quite small again.
Also, the additional leptons in the three-body decays become
energetic enough to pass the lepton cuts, so these events contribute
to the $3\ell$ and $4\ell +$ signals rather than the SS signal.
At the very largest values of $\tan\beta$, $\stilde \tau_1^+ \stilde
\tau^-_1$ becomes by far the dominant discovery process, leading to
essentially only a HIT signal.
Fig.~\ref{varytb} illustrates that the ratio of HITs to SS dimuons
is an interesting probe of $\tan\beta$ in gauge mediated models.

\begin{figure}[th]
\centering
\epsfxsize=4.3in
\hspace*{0in}
\epsffile{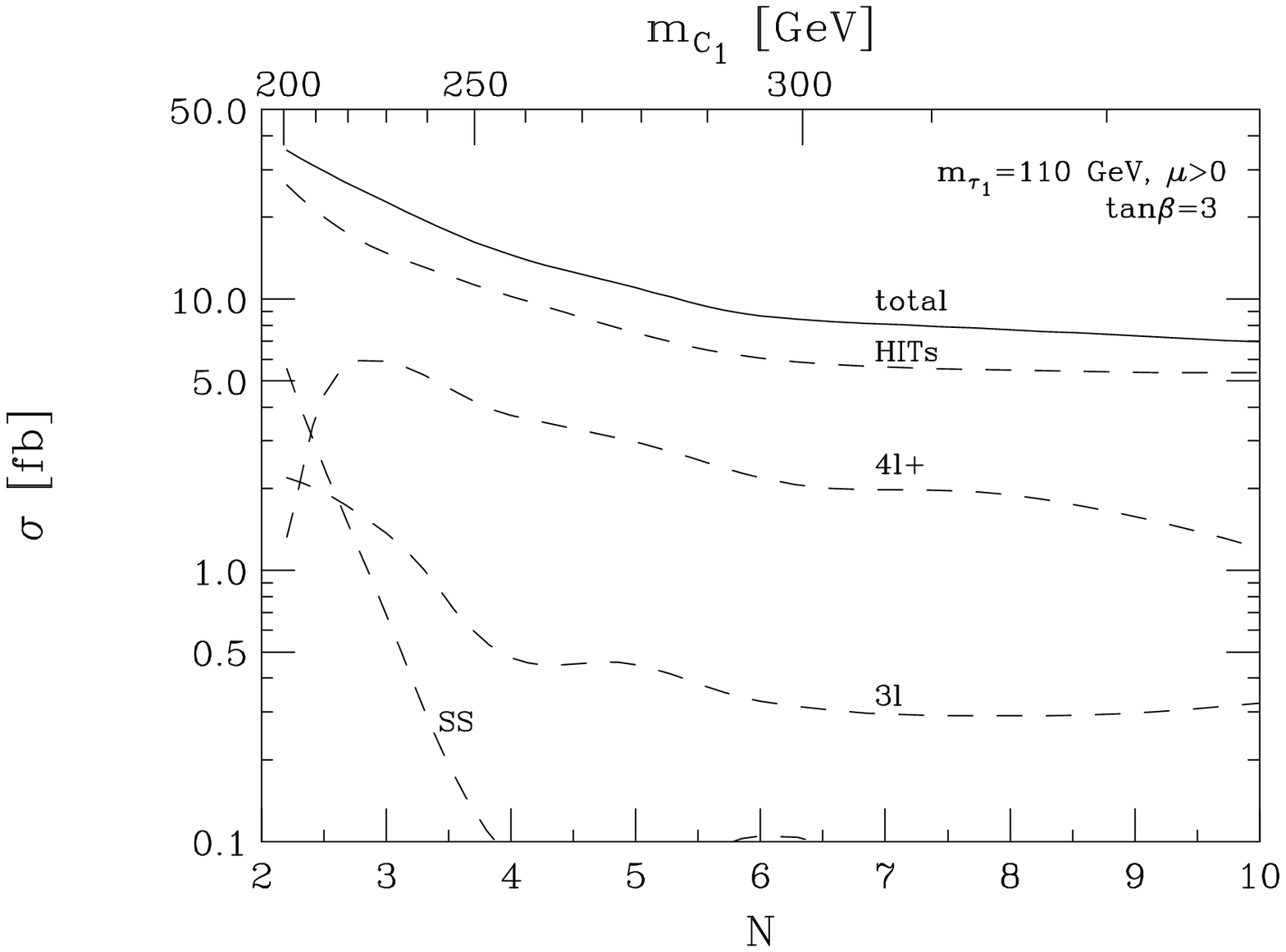}
\caption{As in Figure \ref{tb3varylambda}, but
for slepton co-NLSP models 
with the mass of $\widetilde\tau_1$ fixed at 110 GeV,
$\tan\beta = 3$,
and varying $\Nmess$.
}
\label{tb3varyN}
\vspace*{0.3in}
\epsfxsize=4.3in
\hspace*{0in}
\epsffile{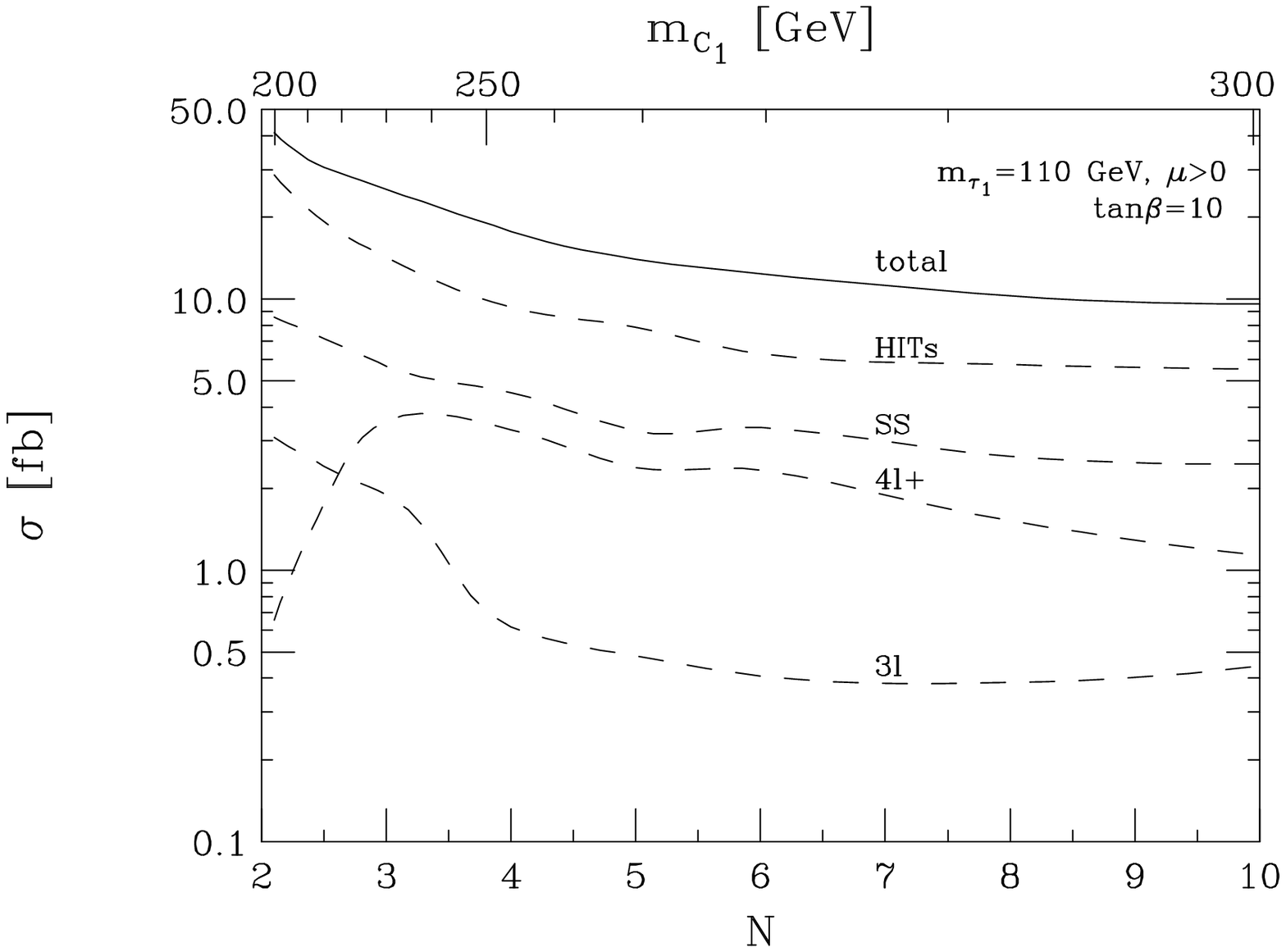}
\caption{As in Figure \ref{tb10varylambda}, but
for stau NLSP models with the mass of $\widetilde\tau_1$ fixed at 110 GeV,
$\tan\beta = 10$,
and varying $\Nmess$.
}
\label{tb10varyN}
\end{figure}

So far we have considered models with fixed $\Nmess=3$.
It is also interesting to consider how the signals change if 
$\Nmess$ is varied,\footnote{In the simplest GMSB
models, $\Nmess$ is taken to be an integer, but one can
easily imagine more general frameworks of models in which the effective
value for $\Nmess$ is not so restricted.}
since this has the effect of changing the overall
ratios of the slepton masses to the gaugino mass parameters.
In
Fig.~\ref{tb3varyN} we show the signals as a function of 
$\Nmess$, 
with $\tan\beta = 3$, $\mu> 0$, and $\Lambda = \Mmess/3$ chosen in such a
way that $m_{\stau_1}$ is held fixed at 110 GeV. 
(The lower endpoint of the graph is determined by the fact that for
$\Nmess \lsim 2.2$, one finds $m_{\stilde N_1} < m_{\stilde e_R},
m_{\stilde \mu_R}$ in these models, so that we are no longer in the
slepton co-NLSP scenario. For smaller $\Nmess$, the NLSP will be
a neutralino, leading to missing energy signals if the decays to
the goldstino/gravitino $\GG$ take place outside the detector.)
For all values of $\Nmess$, the HIT signal is the largest component of
the total.
For the lower values of $\Nmess$, the charginos and neutralinos
are sufficiently light that $\stilde C_1^+ \stilde C_1^-$
and $\stilde C_1^{\pm} \stilde N_2$ production dominate. 
As $\Nmess$ increases,
the $\stilde C_1$ and $\stilde N_2$ decays tend to yield more additional
leptons,
so that the $4\ell+$ signal quickly overtakes the SS signal. For the
largest values
of $\Nmess$, slepton production dominates and there is essentially no
SS signal at all.

The situation is again quite different for stau NLSP models, as
illustrated
in Fig.~\ref{tb10varyN}. Here we have chosen $\tan\beta = 10$
and all other parameters as in Fig.~\ref{tb3varyN}. This ensures
that the mass differences $m_{\stilde e_R} - m_{\stilde \tau_1}$
and $m_{\stilde \mu_R} - m_{\stilde \tau_1}$ are greater than $3\gev$,
so that the three-body decays of $\stilde e_R$ and $\stilde \mu_R$
are open. This in turn guarantees that over the whole range of $\Nmess$
shown, the SS signal is a significant component of the total.  With 
enough integrated luminosity it may be possible to extract information
about the
number of messengers by measuring the superpartner masses and comparing
the $3l$ and $4l+$ rates. Even more information can be obtained by
measuring the proportion of 4, 5, and 6 lepton events in the latter
sample. However, for any given model these ratios are quite strongly
dependent on the choice of lepton $p_T$ cuts which in turn depend on
experimental realities that are difficult to anticipate, so
we will not
analyze them here. 

It is useful to remark on the proportion of $\stilde
e_R^+\stilde e^-_R$ and $\stilde
\mu_R^+\stilde \mu^-_R$ events which lead to same-sign staus in the final
state. For example, in Fig.~\ref{tb10varyN}, 
the ratio of ``slepton charge-flipping" decays
$\stilde \ell_R^- \rightarrow \ell^- \tau^- \stilde \tau_1^+ $
to ``slepton charge-preserving" decays 
$\stilde \ell_R^- \rightarrow \ell^- \tau^+ \stilde \tau_1^- $
increases monotonically from 1 to about 4.6 as $\Nmess$ increases
from the minimum value of
2.1 to 10. This increase is attributable to the corresponding rise in
the ratio $m_{\stilde N_1}/m_{\stilde \ell_R}$, since 
off-shell neutralinos in the three-body decays eq.~(\ref{threebody})
favor the slepton charge-flipping channel, while nearly on-shell
neutralinos do not distinguish between the two channels
\cite{threebodyakm}. This
means that for $\Nmess = 2.1$, nearly 50\% of the
$\stilde
e_R^+\stilde e^-_R$ and $\stilde
\mu_R^+\stilde \mu^-_R$ events will have same-sign staus in the final
state, while
for $\Nmess = 10$, the fraction with same-sign staus decreases to
about 27\%. (Note that for smaller values of $\Nmess$, the deviation
of this fraction from 50\% is much less; for example, in the models
shown in Fig.~\ref{tb10varylambda}, it never gets below about 46\%.)
In addition, there is a large direct 
production of $\stau_1^+ \stau_1^-$ which further dilutes
the ratio of same-sign stau final states. 

It should also be
noted that chargino and neutralino production will lead to some
same-sign events with branching fractions that are 
functions of the model parameters. Certainly, all $\stilde C_1^\pm
\stilde N_2$ events will lead to same-sign quasi-stable sleptons in
the final state, simply because the decay of the neutral Majorana
particle $\stilde N_2$ must be
democratic between different charge channels. In most of the models
studied above, the production cross-section for 
$\stilde C_1^+ \stilde C_1^-$ is larger than for 
$\stilde C_1^\pm \stilde N_2$, but it can also lead to
same-sign sleptons in the final state whenever any part of the decay chain
goes through a real or virtual neutralino. Only a small fraction
of these events will be counted in the SS signal as defined above which
excludes events with HITs or additional leptons. However, it is important
to keep in mind that a sizeable fraction of all the signals will have
same-sign sleptons which can be an important observable. Thus one can,
for example,
measure the charges of the two tracks with the highest $p_T$ which qualify
as a muon candidate or a HIT, and then compare the ratio of same-sign
to opposite-sign charges. This observable can be defined for
each of the signals given above, and should give an
important confirmation of
the slepton interpretation of these events, as well as some information
about the model parameters.

In our study the number
of SS events compared to HITs depends critically on detector performance.
For example, in the limit that detectors cannot distinguish between muons and
heavy charged particles at {\em any} $\beta\gamma$, the HITs signal
will of course go to zero and the other signals will rise.  Ratios
between SS, $3l$ and $4l+$ signals will then have very similar dependences
to those found above, and information about the number of messengers
and $\tan\beta$ can be studied in a similar fashion.  In other words,
the qualitative features do not disappear with variations in the detector
parameters.  In our study we have attempted to mimic 
detector performance similar to that expected at CDF and D0.

In conclusion, quasi-stable
heavy charged particles are present in many extensions beyond the
Standard Model.  As expected the reach in supersymmetry masses is 
very high in this 
scenario since highly-ionizing ``cannonballs" in the detector are hard
to miss.  Mass limits well in excess of the capabilities at LEPII are
possible. Indeed, the Tevatron Run II can probe much of the parameter
space where GMSB sparticles might be expected to appear, based on
a solution to the hierarchy problem without significant fine-tuning
\cite{ibis}.
However, only
a detailed study of relative rates
of new physics final states along the lines of those suggested
above will enable a self-consistent picture
to be formed of gauge-mediated supersymmetry at the Tevatron.

\noindent
{\it Acknowledgments:} 
We are grateful to H.~Baer and D.~Stuart for useful conversations.
The work of SPM was supported in part by the US Department of Energy.


\end{document}